\documentstyle[jkas]{article}
 
\def\etal{{et al.}\ }
  
\runningauthor{ANN ET AL.}
\runningtitle{PHOTOMETRIC SURVEY OF OPEN CLUSTERS}
\beginpage{7}
\endpage{16}

\def\s-1{{\rm\,s^{-1}}}

\def\spose#1{\hbox to 0pt{#1\hss}}
\def\lta{\mathrel{\spose{\lower 3pt\hbox{$\mathchar"218$}}
     \raise 2.0pt\hbox{$\mathchar"13C$}}}
\def\gta{\mathrel{\spose{\lower 3pt\hbox{$\mathchar"218$}}
  \raise 2.0pt\hbox{$\mathchar"13E$}}}

\font\twelvei = cmmi10 scaled\magstep1
\font\teni = cmmi10 
\font\mbf = cmmib10 scaled\magstep1
\font\mbfs = cmmib10 \font\mbfss = cmmib10 scaled 833
\font\msybf = cmbsy10 scaled\magstep1
\font\msybfs = cmbsy10 \font\msybfss = cmbsy10 scaled 833
\def\lsim{\raise0.3ex\hbox{$<$}\kern-0.75em{\lower0.65ex\hbox{$\sim$}}}
\def\gsim{\raise0.3ex\hbox{$>$}\kern-0.75em{\lower0.65ex\hbox{$\sim$}}}

\begin{document}

\title{BOAO PHOTOMETRIC SURVEY OF GALACTIC OPEN CLUSTERS. I. 
BERKELEY 14, COLLINDER 74, BIURAKAN 9, and NGC 2355 }

\author{
{\large\textbf{\textsc{H. B. A}}}\textbf{\textsc{NN}}$^1$, 
{\large\textbf{\textsc{M. G. L}}}\textbf{\textsc{EE}}$^2$, 
{\large\textbf{\textsc{M. Y. C}}}\textbf{\textsc{HUN}}$^3$, 
{\large\textbf{\textsc{S.-L. K}}}\textbf{\textsc{IM}}$^3$, 
{\large\textbf{\textsc{Y.-B. J}}}\textbf{\textsc{EON}}$^3$, 
{\large\textbf{\textsc{B.-G. P}}}\textbf{\textsc{ARK}}$^3$,\\
{\large\textbf{\textsc{I.-S. Y}}}\textbf{\textsc{UK}}$^3$, 
{\large\textbf{\textsc{H. S}}}\textbf{\textsc{UNG}}$^4$, 
\textbf{\textsc{AND}}
{\large\textbf{\textsc{S. H. L}}}\textbf{\textsc{EE}}$^1$
}

\address{\normalsize{$^1$Department of Earth Sciences, Pusan National University, Pusan 609-735}}
\address{\normalsize{$^2$Department of Astronomy, Seoul National University, Seoul 151-742}}
\address{\normalsize{$^3$Bohyunsan Optical Astronomy Observatory, Yeongcheon 770-820}}
\address{\normalsize{$^4$Department of Earth Science, Kyungpook  National 
University, Taegu 702-742}}
\address{\normalsize{\it (Received Mar. 22, 1999; Accepted Apr. 15, 1999)}}

\abstract{
Open clusters are useful tools to investigate the structure and evolution
of the Galactic disk.  We have started a long-term project to obtain $UBVI$ 
CCD photometry of open clusters which were little studied before, using 
the Doyak 1.8 m telescope of Bohyunsan Optical Astronomy Observatory
in Korea.  The primary goals of this project are (1) to make a catalog
of $UBVI$ photometry of open clusters,
(2) to make an atlas of open clusters, 
and (3) to survey and monitor variable stars
in open clusters.  Here we describe this project and report 
the first results based on preliminary analysis of the data 
on four open clusters in the survey sample:  Be~14, Cr~74, Biu~9, and NGC~2355. 
Isochrone fitting of the color-magnitude diagrams of the clusters
shows that all of them are intermediate age to old (0.3 -- 1.6 Gyrs) 
open clusters with moderate metallicity.
}


\maketitle
 
\section{INTRODUCTION}

Open clusters are ideal targets with which to investigate the structure and
evolution of the Galactic disk as well as the formation and evolution of
stars and stellar clusters.  Current understanding of the galactic disk
has been greatly benefited by the studies of open clusters.  Young 
clusters have been used for investigation of spiral arm structures 
(Becker \& Fenkart 1970; Janes \& Adler 1982; Feinstein 1994) and for the study
of the star formation processes in the galactic disk 
(Tarrab 1982; Sagar \etal 1988;
Phelps \& Janes 1993). Old clusters have provided critical informations
about the formation and early evolution of the galactic disk 
(Janes \& Phelps 1994; Friel 1995; Wee \& Lee 1996; 
Twarog, Ashman \& Anthony-Twarog 1997; Lee 1997; Park \& Lee 1999). 
 
There are about 1,200 known open clusters in the Catalogue of
Open Cluster Data (Lynga 1987; hereafter COCD).  In spite of many 
investigations including recent CCD photometries of open clusters, 
little is known
except for the position, angular diameter, and the number of member stars
for the majority of the clusters.  To date only one third of them 
have been studied in detail to derive the basic physical parameters such as
distance, reddening, age, and metallicity.  More than half of them have the
Trumpler richness class $m$ and $r$, while most of the unstudied 
clusters are classified as $p$. 
This shows that the present knowledge of open clusters are heavily biased
by the properties of rich open clusters (Ann 1997).

To understand the nature of clusters and the structure of galactic disk
as a whole, we have initiated a comprehensive photometric 
survey of open clusters which
were little studied before, using the Doyak 1.8m telescope which was newly 
installed at Bohyunsan Optical Astronomy Observatory (BOAO) in Korea. 
The primary goal of the present survey is to obtain accurate color-magnitude diagrams (CMDs) of the clusters with $UBVI$ CCD photometry
and to derive the basic physical parameters of the clusters.
Secondly we will make a BOAO Photometric Atlas of 
Open Clusters which will serve as a useful tool to study the structure
of the Galactic disk and which will be used as a guide to study the 
individual open clusters in detail.  
Along with these main goals, the present survey is designed also
to survey and monitor variable stars in open clusters. The primary 
targets of variable star survey in clusters are short period variable stars
such as $\delta$ Scuti, $\gamma Dor$, $SPB$, and $\beta$ Cephei type stars.

In this paper we describe this survey project in overall and present
the first results of the photometric survey on four open clusters made during 
the observing run of 1998. A brief description of this project was given
by Chun (1998).
In Section II, we describe the project of the photometric survey 
of Galactic open clusters.
The first results of the present survey are given in Section III and  
summary and discussion are given in the last section. 

\section{THE PHOTOMETRIC SURVEY OF OPEN CLUSTERS}

\subsection{The Survey Sample}

First we have selected 343 candidate clusters for the BOAO photometric survey
from COCD by the declination criterion of  $\delta > -20^{\circ}$, which
are observable at the BOAO.
Our sample consists of about half of the clusters with unknown distance
and age in COCD.  Fig.~1 shows the distribution and histogram of the program
clusters. There is almost no cluster in the right ascension
of $8^{h} < \alpha < 16^{h}$ and most of the program clusters can be observed
in the fall and winter season when the weather condition is best at the BOAO.
We list the Trumpler class of the program clusters in Table~1.  As shown in
Table~1, about two thirds of the clusters are classified as richness
class $p$ and about a third as $m$.  The clusters of richness class
 $r$ is less than 5 \% of the sample.  With regard to the concentration class,
about two thirds of the clusters are classified as $III$ or $IV$.  This shows
that most of the unstudied clusters are poor clusters with weak central
concentration.


\begin{table}[t]
\begin{center}
{\bf Table 1.}~~Data of candidate clusters for the BOAO photometric survey\\
\vskip 0.3cm
\begin{tabular}{ccccc}
\hline\hline
{concentration$\backslash$richness} & p & m & r & total\\
\hline
I   &  17 &  18 &   6 &   41 \nl
II  &  38 &  32 &   4 &   74 \nl
III &  51 &  32 &   4 &   95 \nl
IV  &  95 &  22 &   1 &  118 \nl
\hline
total & 201 & 112 &  15&  328 \nl
no data &   &     &    &   15 \nl
\hline
\end{tabular}
\end{center}
\end{table}

\subsection{Observational Strategy}

The target clusters will be observed using the SITe $2048 \times 2048$ CCD 
camera with Johnson-Cousins $UBVI$ filters attached to the F/8 cassegrain 
focus of the Doyak 1.8m telescope. 
The pixel size of the CCD is 24 $\mu\rm{m}$ that
corresponds to 0.$^{\prime \prime}$34 on the sky. The size of the 
covered field of view is $\sim 11.^{\prime}8 \times 11.^{\prime}8$. 
The gain and readout noise of the CCD are 1.8 electrons/ADU and $7$ electrons, respectively.
When the nights are photometric, standard stars (e.g. Landolt 1992) will be
observed for calibration of the photometry as well as the target clusters.
When the nights are non-photometric, variable stars in the target clusters
will be surveyed and monitored.

\section{ THE FIRST RESULTS}

\subsection{Observations}

60 open clusters among the total sample were observed
during the fall observing runs in 1998. We report here the first results
of this project, based on preliminary analysis of the data of 
four open clusters in the sample: Be~14, Cr~4, Biu~9, and NGC~2355.
These clusters were observed 
on the photometric night of 1998, Dec. 16.
We have obtained images of the central $11.^{\prime}8 \times 11.^{\prime}8$ 
regions of each cluster with a pair of exposures, long and short,
for each filter.  Along with the object observations, we have observed
a sufficient number of twilight flat fields before and after the observations
of the objects.
We observed Landolt standard stars in SA~98 (Landolt 1992) together with 
standard stars in M67 (Montgomery, Marschall \& Janes 1993) and
NGC~7790 (Christian 1985) several times during the observations
to transform the instrumental magnitudes to the standard system.

Table~2 shows the observational log of the four clusters. 
 Greyscale maps of the $V$-band images of the clusters 
taken with long exposures are displayed in Fig.~2. 


\begin{table}[h]
\begin{center}
{\bf Table 2.}~~Journal of observations\\
\vskip 0.3cm
\begin{tabular}{ccc}
\hline\hline
Cluster & Filter & T$_{\rm exp}$\\
\hline
 Be 14 & U  &  900 s   \nl
       & B  & $200 s \times 3$, 20 s  \nl
       & V  & $100 s \times 3$, 10 s  \nl
       & I  & $50 s \times 3$, 5 s    \nl
\hline
 Cr 74 & U  & $200 s \times 3$   \nl
 Biu 9 & B  & $200  s \times 3$, 20 s  \nl
 NGC 2355   & V  & $100 s \times 3$, 10 s  \nl
       & I  & $50  s\times 3$, 5  s   \nl
\hline
\end{tabular}
\end{center}
\end{table}

\begin{table}
\begin{center}
{\bf Table 3.}~~Extinction Coefficients and Transformation Coefficients\\
\vskip 0.3cm
\begin{tabular}{ccccc}
\hline\hline
Filter & $k$ & $z$ & $a$ & $\sigma$(fit)\\
\hline
         U  & -0.520 & -4.811 & 0.347 & 0.053 \nl
         B  & -0.306 & -1.919 & 0.186 & 0.035 \nl
         V  & -0.144 & -1.624 & -0.024 & 0.025 \nl
         I  & -0.066 & -1.707 & 0.037 & 0.028 \nl
\hline
\end{tabular}
\end{center}
\end{table}

\subsection{Data Reduction}

We have followed the standard technique of CCD reductions 
using IRAF/{\small CCDRED}
for basic reductions of the obtained images.  
This includes the subtraction of bias frames with
overscan correction, trimming, and flatfielding.  
For stellar photometry, IRAF/DAOPHOT 
were used
to estimate the instrumental magnitudes of the cluster stars and standard
stars, respectively.  We applied aperture corrections to the instrumental 
magnitudes of cluster stars obtained from PSF fitting, 
using the instrumental magnitudes obtained with an aperture radius 
of $7^{\prime \prime}$.
 

The transformation of the instrumental magnitudes to the standard system 
was made by two steps.  First, we applied the atmospheric extinction
correction by the primary extinction coefficient derived from the multiple 
observations of M~67 and NGC~7790.  Second, we transform the extinction
corrected instrumental magnitudes to the magnitudes in the standard system,
by the following transformation equations 
$$U = u - z_{u}  - k_{u}  X - a_{u} (U-B) $$
$$B = b - z_{b}  - k_{b}  X - a_{b} (B-V) $$
$$V = v - z_{v}  - k_{v}  X - a_{v} (B-V) $$
$$I = i - z_{i}  - k_{i}  X - a_{i} (V-I) $$
where the capital letters stand for the magnitudes in the standard system,
the lower case letters for the instrumental magnitudes, and $X$ for the
airmass.  The transformation coefficients $z$'s and $a$'s are 
determined by a least-squares fit of the instrumental magnitudes of the 
Landolt standard stars to the magnitudes in the standard system (Landolt 1992),
after correction of the atmospheric extinction.  
Fig.~3 displays the residuals of the calibration of the present photometry,
showing that the standard deviations of the calibration are less than
0.04 mag except for $U$ which has a standard deviation of 0.05 mag. 
Table~3 lists the resulting coefficients for the
transformation equations.

\subsection{Results}

We have obtained the color-magnitude and color-color diagrams of Be~14,
Cr~74, Biu~9, and NGC~2355 from the present photometry. 
All stars in the cluster images, with photometric errors
less than 0.1 mag, are used to construct the cluster CMDs but only the
stars with $V < 17$ are used for the color-color diagrams.  
The distance, reddening, age, and metallicity of the observed clusters 
are determined simultaneously by fitting the Padova isochrones (Bertelli 
\etal 1994) to the observed stellar distributions in CMDs of the clusters.
We have also obtained independent estimates of the reddenings of the clusters 
from the observed stellar distributions in $(U-B)$--$(B-V)$ diagrams
(Schmidt-Kaler 1982), which
are used as a guide to isochrone fitting.
We assumed a slope of $E(B-V)/E(U-B)$ = 0.72,
 the total to selective extinction ratio ${R_V}$=3.0, and
$E(V-I)=1.25 E(B-V)$ for the analysis of the data 
in this study (Johnson \& Morgan 1953; Dean, Warren \& Cousins 1978).
In the isochrone fittings, we put the most weight on the
stars in the turnoff regions and giant branches, especially 
the stars located within the cluster diameter, represented
by open circles  in Figs.~4 -- 7.  
For the main sequence region, we tried to fit the lower envelope of the main
sequence, considering the photometric errors.  

\subsubsection{Berkeley 14}

Be~14 is located in the constellation Auriga ($\alpha_{2000}=5^{h} 0^{m}$,
$\delta_{2000}=43^{\circ} 28.^{\prime}6$, $l=162^{\circ} 52^{\prime}$, 
$b=0^{\circ} 43^{\prime}$). 
 The angular diameter of the cluster in COCD is $5^{\prime}$.
The Trumpler class of the cluster is $III1m$, showing that it is detachable
from field stars but has little concentration. 
However, it is 
difficult to isolate the cluster from the background field stars due to 
heavy contamination of field stars in the galactic plane.
Fig.~4 shows the color-color diagram with fiducial
ZAMS lines together and the CMDs of the measured stars in Be 14.
It is seen in the CMDs that Be 14 shows 
a typical feature of old open clusters: 
a red giant branch with well-distinguishable
red giant clump at $V\approx 15.5$ mag. 
A main sequence turnoff of Be 14 is seen at $V\approx 17$ mag.    
The best fitting isochrone of log(t[yrs]) = 9.2 
and [Fe/H] $= -0.32$ dex, with reddening of $E(B-V)=0.52$ and
a true distance modulus of $(m-M)_{0}=13.7$ is overlayed on the CMDs.  
As shown in Fig.~4, the fittings of the isochrone to the stellar 
distributions in the CMDs of Be~14 are quite good, especially for the 
stellar distributions in the base of giant branch and the clump stars located
within the cluster diameter, designated as open circles. 

\subsubsection{Collinder 74}

Cr 74 is located in the west 
of $\alpha$ Ori ($\alpha_{2000}=5^{h} 48^{\prime} 30^{\prime \prime}$,
$\delta_{2000}=7^{\circ} 24.^{\prime}0$, $l=198^{\circ} 59^{\prime}$,
$b=-10^{\circ} 24^{\prime}$), in the direction of the anticenter of our Galaxy.
The angular diameter of Cr 74 given in the COCD is 
about $5^{\prime}$, and the Trumpler class of this cluster is
$III1m$.  As inferred from the same Trumpler class as Be~14, 
the appearance of the cluster is similar to that of Be~14, but it is 
more difficult to isolate Cr~74 from the field stars than 
Be~14. 

Isochrone fittings in Fig.~5 show a pretty good match of the isochrone
of log(t[yrs])=9.1 and     
[Fe/H] = +0.07 dex with $E(B-V)$ = 0.38 and $(m-M)_{0}=12.0$
to the observed stellar distributions of Cr~74.  
Although the lower main sequence is heavily contaminated by field stars,
the good match of the isochrone to the stars in the turnoff region is 
remarkable. 

\subsubsection{Biurakan 9}
Biu~9 is located in the constellation
Monoceros ($\alpha_{2000}=6^{h} 57^{\prime} 43^{\prime \prime}$,
$\delta_{2000}=3^{\circ} 13.^{\prime}0$, $l=210^{\circ} 48^{\prime}$,
$b=2^{\circ} 53^{\prime}$).  As shown in Fig.~2, it is easy to identify 
the cluster because of the slight concentration of
stars toward the center with moderate number of stars. This property is 
consistent with the Trumpler class of $II2m$. 


Fig.~6 shows the color-color diagrams and CMDs of Biu~9 with ZAMS
fitting and isochrone fittings, respectively. The field stars contaminate
the photometric diagrams heavily, but they are well separated in the
color-color diagrams by two groups of stars: 
one group with blue $B-V$ colors and red $U-B$ colors, and the other
with red $B-V$ colors and blue $U-B$ colors.   
\begin{samepage}
We have derived
the physical parameters of Biu~9 from the stars with blue $B-V$ colors,
because the stars with red $B-V$ colors could not be matched by the fiducial 
ZAMS and isochrones, simultaneously.  
As shown in Fig.~6, the isochrone of
log(t[yrs]) = 8.5 and [Fe/H] $=-0.32$ dex  
with $E(B-V)$ = 0.50 and $(V-M_{V})_{0}=13.4$ 
well matches the stellar distributions in CMDs of Biu~9. 

\begin{table*}
\begin{center}
{\bf Table 4.}~~Basic parameters of the four open clusters\\
\vskip 0.3cm
\begin{tabular}{ccccccc}
\hline\hline
Cluster & Angular Diameter & $E(B-V)$ & $(m-M)_0$ & 
$R_{\rm GC}$ & [Fe/H] & Age \\
\hline
Be 14 & 5$^{\prime}$ & $0.52\pm0.04$ & $13.7\pm0.2$ & 13.8 kpc & $-0.32$ dex & $1.6\pm0.2$ Gyrs \nl 
Cr 74 & 5$^{\prime}$ & $0.38\pm0.04$ & $12.0\pm0.2$ & 10.9 kpc & $+0.07$ dex & $1.3\pm0.2$ Gyrs \nl 
Biu 9 & 3$^{\prime}$ & $0.50\pm0.04$ & $13.4\pm0.2$ & 12.8 kpc & $-0.32$ dex & $0.3\pm0.04$ Gyrs \nl 
NGC 2355 & 7$^{\prime}$ & $0.25\pm0.02$ & $11.4\pm0.1$ & 10.3 kpc & $-0.32$ dex & $1.0\pm0.1$ Gyrs \nl 
\hline
\end{tabular}
\end{center}
\end{table*}

\subsubsection{NGC~2355}

NGC~2355 is located in the constellation of 
\end{samepage}
Gemini ($\alpha_{2000}=7^{h} 16^{\prime} 55^{\prime \prime}$,
$\delta_{2000}=13^{\circ} 46.^{\prime}7$, $l=203^{\circ} 22^{\prime}$,
$b=11^{\circ} 48^{\prime}$).  The Trumpler class of the cluster, $II2m$
is the same as that of Biu~9 but it shows weaker central concentration
than Biu~9. However, it is not difficult to identify the cluster 
in the CCD images. 

The color-color diagram and CMDs of NGC~2355 in Fig.~7 show a well defined 
main sequence with a large number of field stars.  We applied isochrone
fittings to derive the physical parameters of NGC~2355, together with the 
ZAMS fitting for reddening determination.  As shown in Fig.~6, the isochrone
of log(t[yrs]) = 9.0 and [Fe/H] $=-0.32$ dex with $E(B-V)$=0.25 
and $(m-M)_{0}=11.4$ is well fitted to the main sequence of NGC~2355
as well as the stars in the turnoff region in $V$--$(B-V)$ diagram.
However, there are some discrepancies between the isochrone and the
stellar distributions in $V$--$(V-I)$ diagram. The reason for this mismatch is
unclear but it might be caused by large photometric errors due to 
non-photometric condition at the time of observing NGC~2355. 
There seems to be a main
sequence gap similar to that observed in NGC~2420 (Anthony-Twarog \etal 1990;
Lee, Ann, \& Kang 1998).  The present estimate of the age of NGC~2355 is 
similar to that obtained by Colegrove \etal (1993). 

\section{DISCUSSION AND SUMMARY}

We have described the BOAO photometric survey of galactic open clusters.
The primary goal of this survey is to
derive physical parameters of the target open clusters using $UBVI$ CCD 
photometry obtained at 1.8m telescope of the BOAO.
The total sample of the BOAO photomeric survey consists of
343 open clusters with unknown distance and ages, selected from COCD by
the declination criteria of $\delta > -20^{\circ}$. 
We present the first results of this
survey, based on preliminary analysis of the data on four open clusters:
Be~14, Cr~74, Biu~9, and NGC~2355, which are subsamples
of 60 open clusters that were observed in the first year of the survey.

We have determined the distance, reddening, age, and metallicity of the
four clusters by isochrone fittings of the CMDs which are obtained from $UBVI$ 
CCD photometry. Table 4 lists a summary of the basic parameters of the
four clusters determined in this study.
 The reddening estimates of the clusters by 
matching the fiducial ZAMS to the stellar distributions in color-color diagrams 
of the clusters are used as constraints for the isochrone fittings.  
The resulting physical parameters of the clusters show that all of the present 
sample of clusters are intermediate age to old 
clusters with moderate metallicity.

The CMDs of the four clusters show morphological 
properties of intermediate age open clusters, which are characterized by 
a continuous sequence of stars from main sequence to giant branch with a
clump of stars in the red giant branch except for the youngest open cluster
Biu~9.  The oldest cluster Be~14 shows a large number of clump
stars with well developed giant branch.  The well defined turnoff regions of
Cr~74, Biu~9, and NGC~2355 allow us reliable isochrone fitting.
The main sequence gap near turnoff, which is a characteristics of intermediate
age open clusters, seems to be present in NGC~2355 at $M_{V} \sim 2$ mag.  
Because the gap is thought to be caused by the hydrogen core 
exhaustion phase (McClure,Forrester, \& Gibson 1974;
Demarque, Sarajedini, \& Guo 1994), we expect
a similar gap in Cr~74 whose age is similar to that of NGC 2355.
Fig. 5 shows that there is a hint of the gap near the main sequence
turnoff of Cr 74, but the presence of the gap in Cr~74 is not 
as evident as that of NGC~2355.  

The analysis of the present photometry indicates that our photometry is 
accurate enough to derive the physical parameters of open clusters observed 
on photometric nights at the BOAO. 
When this survey is completed, the resulting data will serve as a fundamental
database for investigation of the structure and evolution of the
galactic disk as well as a guide for the study of individual open clusters
in detail.  

\acknowledgements
This work was supported in part by the Ministry of Science and Technology
research fund, and
in part by the Basic Science Research Institute Program, Ministry of Education, 
1998, BSRI-98-5411.

\end{document}